\documentclass[aps,preprint,twocolumn,byrevtex,10pt]{revtex4}%
\usepackage{amsfonts}
\usepackage{amsmath}
\usepackage{amssymb}
\usepackage[tcidvi]{graphicx}%
\begin{document}
\preprint{ }
\title[ ]{A quench-cooling procedure compared with the gate-protocol for aging
experiments in the Electron-Glass}
\author{Z. Ovadyahu}
\affiliation{The Hebrew University, Jerusalem 91904, Israel }

\begin{abstract}
Anderson-insulating indium-oxide films excited far from equilibrium exhibit a
variety of memory effects including aging. Full-aging has been recently
demonstrated in this system using two different experimental protocols. The
first, (gate-protocol) employed a MOSFET structure and involved switching
between two gate voltages. In a different procedure, the system was subjected
to a non-ohmic longitudinal field $F$\ for a waiting-time $t_{w}$, and the
relaxation of $G$ was monitored after the field was switched back to its
linear response value. In this paper we describe yet another protocol that
involves measuring the response of the system that has been 'aged' at some low
temperature $T_{L}$ for a duration $t_{w}$ after it was quench-cooled from
high temperature $T_{H}$. Like in the previous protocols, this procedure
results in full-aging behavior. The advantages and shortcomings of the
quench-cooling protocol are pointed out. The results of aging experiments
based on the better-controlled, gate-protocol performed with different systems
are compared and discussed.


\end{abstract}
\maketitle

\section{ Introduction}

The combination of sufficiently strong disorder and interactions in a
degenerate Fermi system may precipitate a glassy state in an electronic
system. Such a scenario, referring to the state as `electron glass', was
offered in several theoretical papers two decades ago \cite{1,2,3}. Further
elaboration of the early models, including quantum effects, suggested that the
glassy phase occurs on the insulating side of the metal-insulator transition
\cite{4}. Experimental evidence for glassy effects in disordered electronic
systems were reported in few cases \cite{5,6,7}. The experiments showed
several transport features that are typical of other glasses, e.g., slow
relaxation and memory effects including aging. Aging is an intriguing memory
effect common to many glassy systems \cite{8}. The system is said to exhibit
aging if the response (e.g., relaxation from some excitation) depends on the
system history in addition to the time $t$. That is in contrast to ergodic
systems where the response depends only on $t$. In electron glasses, this
phenomenon may be observed in the\ relaxation of the excess conductance
$\Delta G(t)$ created by changing the gate voltage from a state at which it
equilibrated for $t_{w}$ back to a state where it was at full equilibrium
(employing a MOSFET structure) \cite{9,10}. This will be referred to here as a
`gate-protocol'. It turns out that the aging function $\Delta G(t,t_{w})$ in
this case is just a function $f$ of $t/t_{w}~$\cite{9,10}. This so called
`full' aging behavior has been rarely observed in such a clean form in any
other glassy system. Recently, another aging protocol was employed in
Anderson-localized indium oxide films by "stressing" the sample with a large
electric field during $t_{w}$ (`stress-aging-protocol') \cite{11}. The latter
procedure is fundamentally different than the gate-protocol in that during the
waiting time the system is \textit{excited} as opposed to \textit{relaxed} in
the gate-protocol. Full aging was realized in this protocol as well. Although
the relaxation laws of the two protocols were not identical, both show a
$log(t)$ dependence for $t\ll t_{w}$ and a slower relaxation for $t\geq t_{w}%
$. This means that the signature of $t_{w}$ is imprinted on the form of each
relaxation curve (in addition of course to the its effect on the
\textit{amplitude} of $\Delta G,$ which is a necessary ingredient for the
aging phenomenon to hold).

In this paper we describe yet another protocol to observe aging in an electron
glass. Motivated by recent theoretical work that shows aging in the electron
glass \cite{12}, the protocol involves quench-cooling the sample from a high
temperature $T_{H}$ to low temperature $T_{L}\ll T_{H}$, and letting it
partially relax for $t_{w}$. The aging is tested by re-exciting the sample
with a sudden change of the gate voltage, and the relaxation of the excess
conductance $\Delta G(t)$ is measured from this time onward (`T-protocol').
Several weaknesses, inherent in the T-protocol are pointed out. Qualitatively
however, the results are similar to these obtained with the gate-protocol and
exhibit full-aging behavior; the ensuing relaxation $G(t,t_{w})$ could be cast
as $f(t/t_{w})$ over a wide range of $t_{w}$. Finally, we compare the aging
function obtained with the more reliable gate-protocol with the trap model
suggested by Bouchaud and Dean. The implications for models of aging in
quantum glasses \cite{4}, where dynamics in phase space is controlled by
tunneling rather than by thermal activation, are discussed.

\section{Experimental}

\subsection{Sample preparation and measurement techniques}

Several different batches of samples were used in the present study. All were
thin films (50$\pm2$\AA \ thick) of crystalline indium-oxide that were e-gun
evaporated on 110$\mu$m cover glass. Sample size was typically 1x1 mm and
their sheet resistance R$_{\square}$ were in the range 1.5-55 M$\Omega$ at
$4.11K$. Gold film ($\approx$1000\AA \ thick) was evaporated on the backside
of the glass and served as the gate electrode. Conductivity of the samples was
measured using a two terminal ac technique employing a 1211-ITHACO current
pre-amplifier and a PAR-124A lock-in amplifier. Except when otherwise noted,
the measurements reported here were performed with the samples immersed in
liquid helium at $T=4.11K$ held by a 100 liters storage-dewar, which allowed
long term measurements of samples as well as a convenient way to maintain a
stable temperature bath. The ac voltage bias was small enough to ensure linear
response conditions (judged by deviation form Ohm's law being within the
experimental error). Fuller details of sample preparation, characterization,
and measurements techniques are given elsewhere \cite{13}.

\subsection{Description of the aging protocols}

The T-protocol used in the experiments described below involved the following
steps. First, the sample, attached to a measuring probe along with a
calibrated $Ge$ thermometer, is raised above the $^{4}$He liquid level. The
conductance of the sample and the reading of the $Ge$ thermometer are
continuously monitored in the process of warming the sample to a temperature
$T_{H}$. During the time the sample dwells at $T_{H}$ the gate voltage $V_{g}$
is set to an initial value, usually $V_{g}^{o}=0$. Care was taken to ensure
that the sample is always in helium atmosphere to minimize irreversible
changes of the sample during temperature cycling. With this setup, warming up
the sample and settling its temperature at $T_{H}$ usually took 3-10 minutes.
The reverse operation, cooling the sample from $T_{H}$ to $4.11K$ could be
accomplished in 3-4 seconds. As will become clear below, this stage of the
protocol is critically important for an aging experiment of the T-protocol
type. The sample conductance $G$ was then plotted as function of the time that
elapsed from the moment the $Ge$ thermometer reached $4.11K$. Such a plot is
shown in figure 1 for a typical run.
\begin{figure}
[ptb]
\begin{center}
\includegraphics[
trim=0.000000in 4.363323in 0.000000in 1.248331in,
natheight=14.583300in,
natwidth=9.791400in,
height=2.917in,
width=3.1808in
]%
{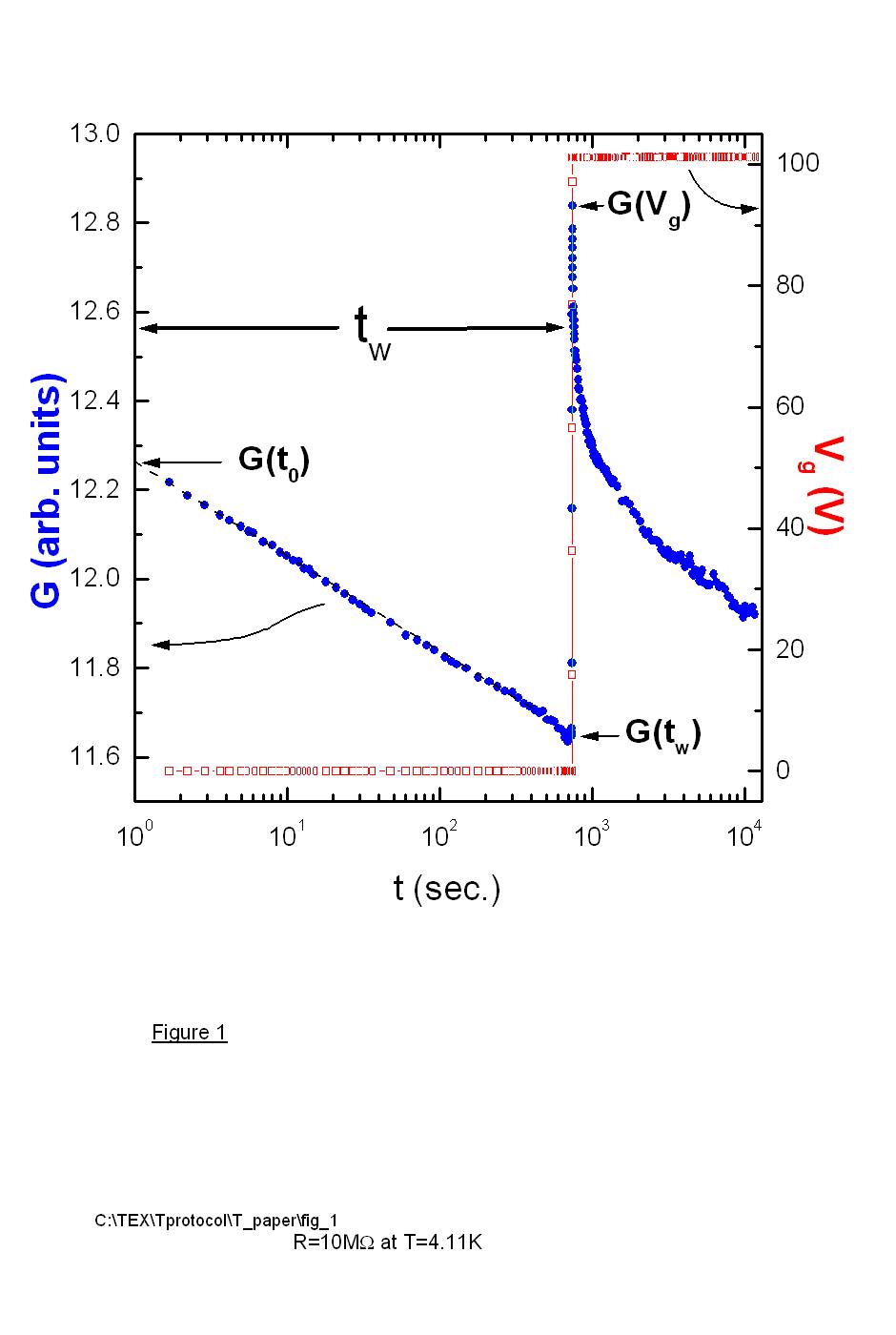}%
\caption{A typical run of the T-protocol. The data marked by full circles is
the sample conductance (left y-axis), and the squares stand for the
corresponding values of the gate voltage (right y-axis). Note the initial
logarithmic relaxation, lasting in this case over a waiting time
$t_{w}=745\sec$. Sample resistance at $T=4.11K$ is $10.4M\Omega.$}%
\end{center}
\end{figure}
Note that, after the quench, $G$ decreases logarithmically. This is the
natural (`history-free') relaxation law of the electron glass; as will be
shown below it may persist for almost six decades in time. In this protocol,
however, the sample is allowed to relax only for $t_{w}$ at which time the
gate voltage is quickly changed (typically over 5 seconds) from $V_{g}^{o}$ to
$V_{g}^{n}$. This causes $G$ to increase, and then, while $V_{g}$ is held
constant, it relaxes again towards a new equilibrium value set by $V_{g}^{n}$
and the bath temperature (c.f., figure 1). Finally, the time dependence of $G$
is recorded for a time span that extends at least three times $t_{w}$. The
same procedure is then repeated with exactly the same set of parameters
($T_{H}$ ,$V_{g}^{o}$ ,$V_{g}^{n}$ , and the changeover time of $V_{g}^{o}$ to
$V_{g}^{n}$ ) except by using each time a different value for $t_{w}$. The set
of plots $G(t,t_{w}),$ where $t$ is measured with respect to the time at which
$V_{g}$ settled at $V_{g}^{n}$, is used as the data for the analysis of the
aging behavior as detailed in the next section.

The gate protocol (more fully described e.g., in reference 9) starts after the
sample equilibrated at $T_{L}$ under $V_{g}^{o}.$ Then, $V_{g}$ is switched to
$V_{g}^{n}$ and allowed to relax for $t_{w}.$ Finally, $V_{g}^{o}$ is
re-instated, and the excess conductance that ensues (relative to the
equilibrium value of $G$) is plotted versus time $t$. A constant bath
temperature is maintained for the entire process.

\section{Results and discussion}

The results reported here were reproduced on two different batches of samples.
In the following however, we describe in detail the results on a single sample
where extensive measurements were done, and where the conductance at a given
temperature was least affected by the repeated temperature cycling. For the
sample shown here the variance of $G$ at $T=4.11K$ was within 1\% for the
entire series involving the cycling between $T_{H}$ and $4.11K$. This is an
impressive figure considering that, for the sample used here, $G$ changes by
two orders of magnitude in this temperature range. The set of data that
resulted from using the T-protocol with six different waiting times and with
$T_{H}\approx100K\pm5K$ is plotted in figure 2.
\begin{figure}
[ptb]
\begin{center}
\includegraphics[
trim=0.000000in 1.246872in 0.000000in 0.624165in,
natheight=14.583300in,
natwidth=9.791400in,
height=4.1208in,
width=3.1808in
]%
{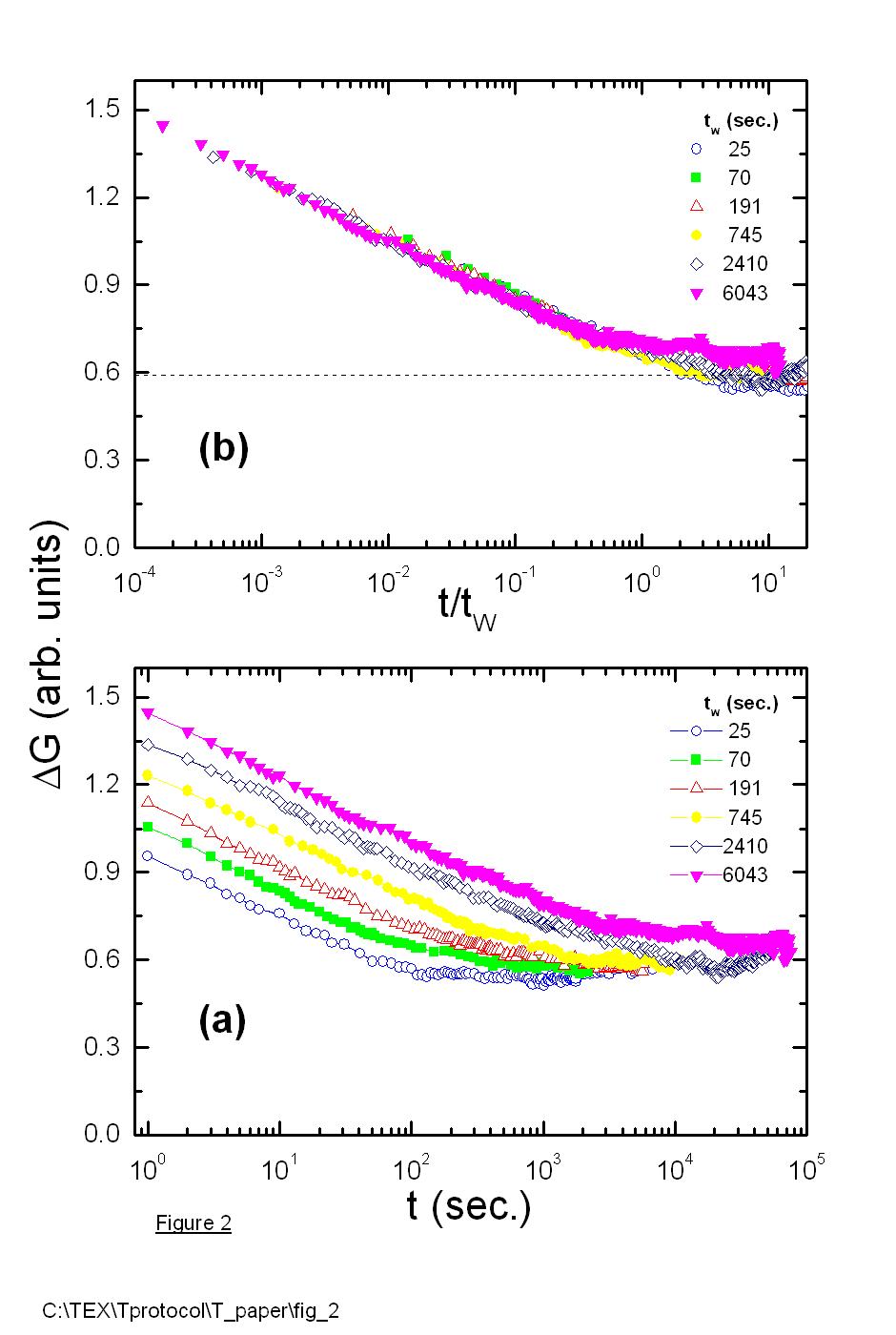}%
\caption{The excess conductance $\Delta G$ (see text) as function of time (a),
and time in units of $t/t_{w}$ (b) for various waiting times. Same sample as
in figure 1.}%
\end{center}
\end{figure}
The quantity $\Delta G(t)$ plotted in the figure is the excess conductance
affected by switching of $V_{g}^{o}$ to $V_{g}^{n}$. Namely, $\Delta G(t)$ is
$G(t)-G_{b}$ where $G(t)$ is the conductance versus time starting from the
moment $V_{g}$ attained the target value $V_{g}^{n},$ and $G_{b}$ is the
`background conductance', a time dependent quantity reflecting the relaxation
of $G$ following the quench. This subtraction also compensates for the
$\pm1\%$ variation in the value of $G$ immediately after a quench in different
runs. $G_{b}$ is a function of both time and $V_{g}^{o}$ and is obtained by
extrapolating towards the end-point of each run along the $\log(t)$ curve
defined by the $t_{w}$ relaxation (the curve recorded during the time interval
marked by $G(t_{0})$ and $G(t_{w})$ in figure 1, for example).

Before discussing the results, the reasons for choosing the various
parameters, in particular the $100K$ for $T_{H},$ should be mentioned. A low
temperature for $T_{H}$ is beneficial in reducing the quench time and the risk
of irreversible sample change due to the thermal cycling. There is however a
concern that favors a relatively high $T_{H}$ in this experiments, and it is
of a rather fundamental nature. In addition to slow relaxation, a common
feature of all glasses is a slow approach to a steady state; For example,
applying a large voltage (field) across an electron-glass results in a time
dependent increase of the conductance, a process that may last for days
\cite{11,14}. The same behavior is observed when the bath temperature is
changed from $T_{1}$ to $T_{2}>T_{1},$ and the effect is actually quite
prominent as illustrated in figure 3 for a specific case.
\begin{figure}
[ptb]
\begin{center}
\includegraphics[
trim=0.000000in 3.740616in 0.000000in 1.248331in,
natheight=14.583300in,
natwidth=9.791400in,
height=3.1168in,
width=3.1808in
]%
{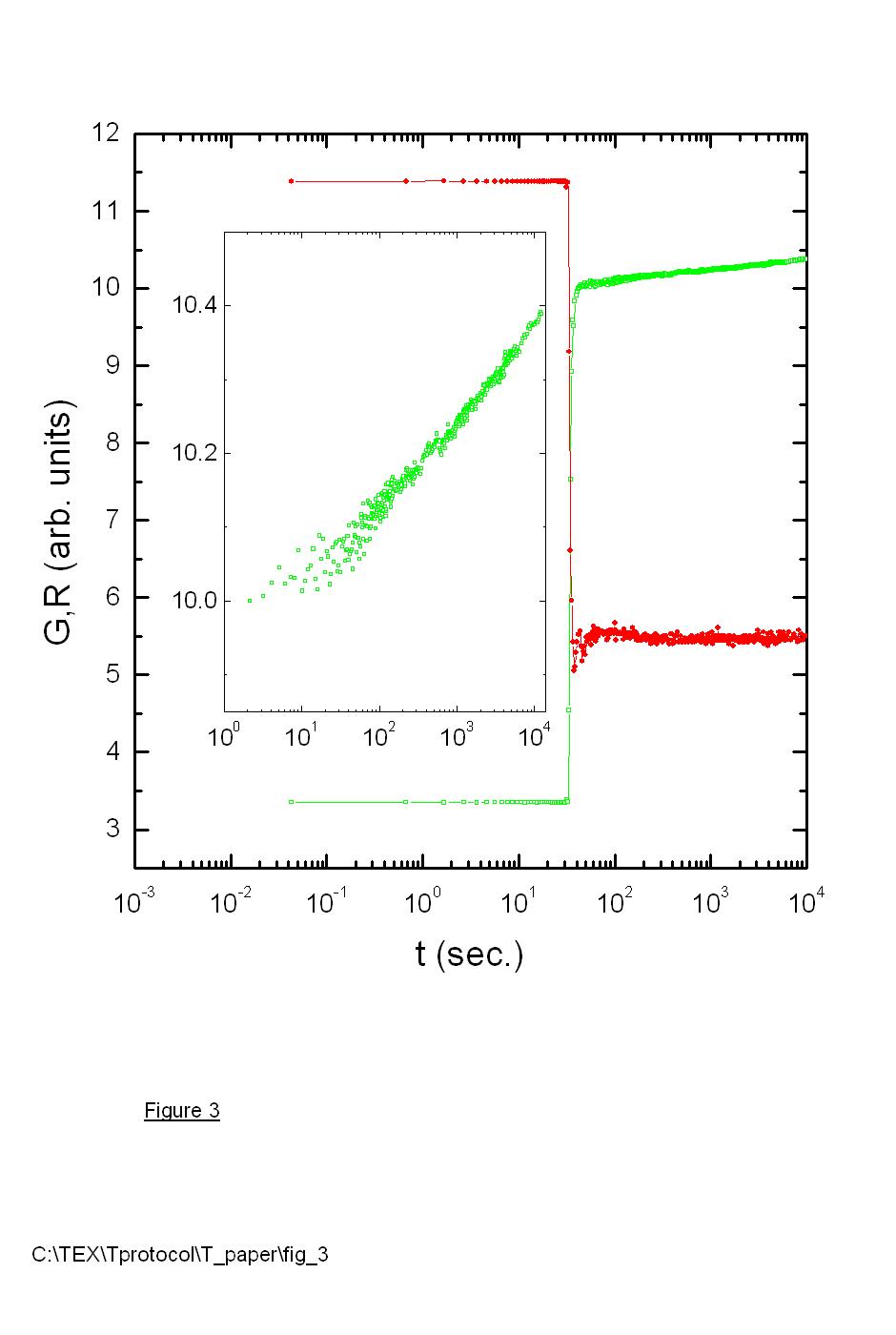}%
\caption{The sample conductance $G$ (squares) and the Ge thermometer
resistance $R$ (circles) during heating (started at $t\approx35\sec.)$ to
$T=6.22K$ from an equilibrated state at $T=4.11K~$and at later times, when the
system temperature has reached a constant value (for $t\eqslantgtr50\sec$.)$.$
Note that $G$ continues to increase logarithmically long after the temperature
has stabilized as judged by the resistance versus time reading of the Ge
thermometer (inset). The sample resistance is $1.42M\Omega$ at $T=4.11K.$}%
\end{center}
\end{figure}
Clearly, this is a complicating feature that introduces additional parameters
to control in an already involved procedure \cite{15}. Fortunately, these (and
other) non-ergodic effects become exponentially smaller with temperature
\cite{16} and they are practically absent in our samples at $T\geq90-100K$.
Using these temperatures for $T_{H}$ instantly removed all traces of memory
from the system, which made the ensuing results independent of the dwell time
at $T_{H}$.

It is essential that the quench-cooling process will be fairly fast and
smooth. If during the cooling process the sample is allowed to spend time at
an intermediate temperature $T_{i}$ (in the range where non-ergodic effects
are noticeable), a signature of this $T_{i}$ will appear at the ensuing
relaxation. This is one of the weaknesses of all protocols based on quench
cooling. Another problem that may actually become more severe when the cooling
process is fast is the appearance of extra noise in the measurement after the
quench. This takes the form of sudden bursts of fluctuations, with
intermittent periods of activity that subsides after few hours. The problem
presumably results from release of mechanical energy due to strains created in
the process of quench-cooling the substrate/sample structure. Repeated thermal
cycling from temperatures somewhat higher than the intended $T_{H}$
('training') usually helped to minimize this disturbance.

The other important parameter in the experiment is $\delta V_{g}=|V_{g}%
^{n}-V_{g}^{o}|$ for which we chose $100V$ as a value large enough to have a
sizeable re-excitation and small enough not to destroy the memory of the
system \cite{13,17}.

Figure 4 shows the respective outcome of the gate-protocol for the same sample
for comparison.
\begin{figure}
[ptb]
\begin{center}
\includegraphics[
trim=0.000000in 1.246872in 0.000000in 0.624165in,
natheight=14.583300in,
natwidth=9.791400in,
height=4.1208in,
width=3.1808in
]%
{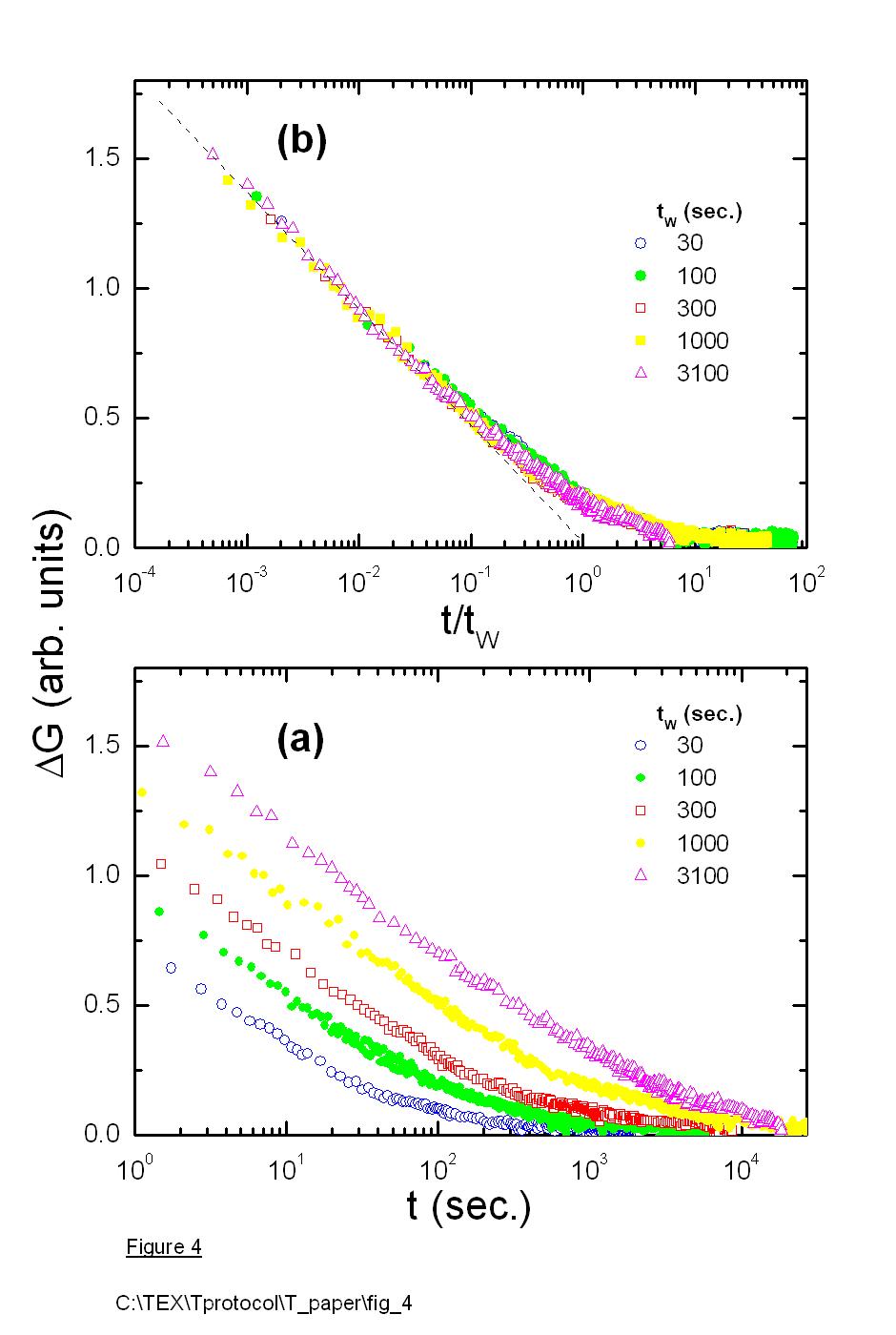}%
\caption{Aging\ experiment using the gate-protocol with five identical values
of waiting-times t$_{w}$. The same values of V$_{g}^{o}$ and V$_{g}^{n}$ used
in the T-protocol of figure 2 were employed. Namely, $V_{g}^{o}=0V,$ and
$V_{g}^{n}=100V.$ Same sample as in figures 1 and 2.}%
\end{center}
\end{figure}
The results of the two aging protocols exhibit several common features. In
both cases, the initial amplitude increases with $t_{w}$, and the ensuing
relaxation is more sluggish at any given $t$ (compare figure 2a and 4a). Also,
upon normalizing $\Delta G$ by $t_{w}$ measured in each case, the curves
collapse onto a common master plot (figures 2b and 4b), and both master plots
exhibit -$\log(t)$ for $t\ll t_{w}$ and a slower dependence for $t\geq t_{w}.$
A notable difference is the non-zero asymptotic value of $\Delta G(t/t_{w})$
in the T-protocol (figure 2b). This however is merely the excess conductance
associated with the equilibrium field-effect, which is due to changing
$V_{g}^{o}$ to $V_{g}^{n}$ (a positive change in this particular case). This
can be appreciated with the help of figure 5 that shows the conductance versus
gate voltage for two stages of the aging process. The figure shows the `memory
cusp', which is a specific modulation in the conductance versus gate-voltage
plot $G(V_{g})$. It is seen by sweeping the gate voltage over a range
including $V_{g}^{o}$ where the system is equilibrating. As noted elsewhere
\cite{13}, the modulation that constitutes the cusp evolves with time during
the equilibration process.
\begin{figure}
[ptbptb]
\begin{center}
\includegraphics[
trim=0.000000in 3.740616in 0.000000in 1.871037in,
natheight=14.583300in,
natwidth=9.791400in,
height=2.917in,
width=3.1808in
]%
{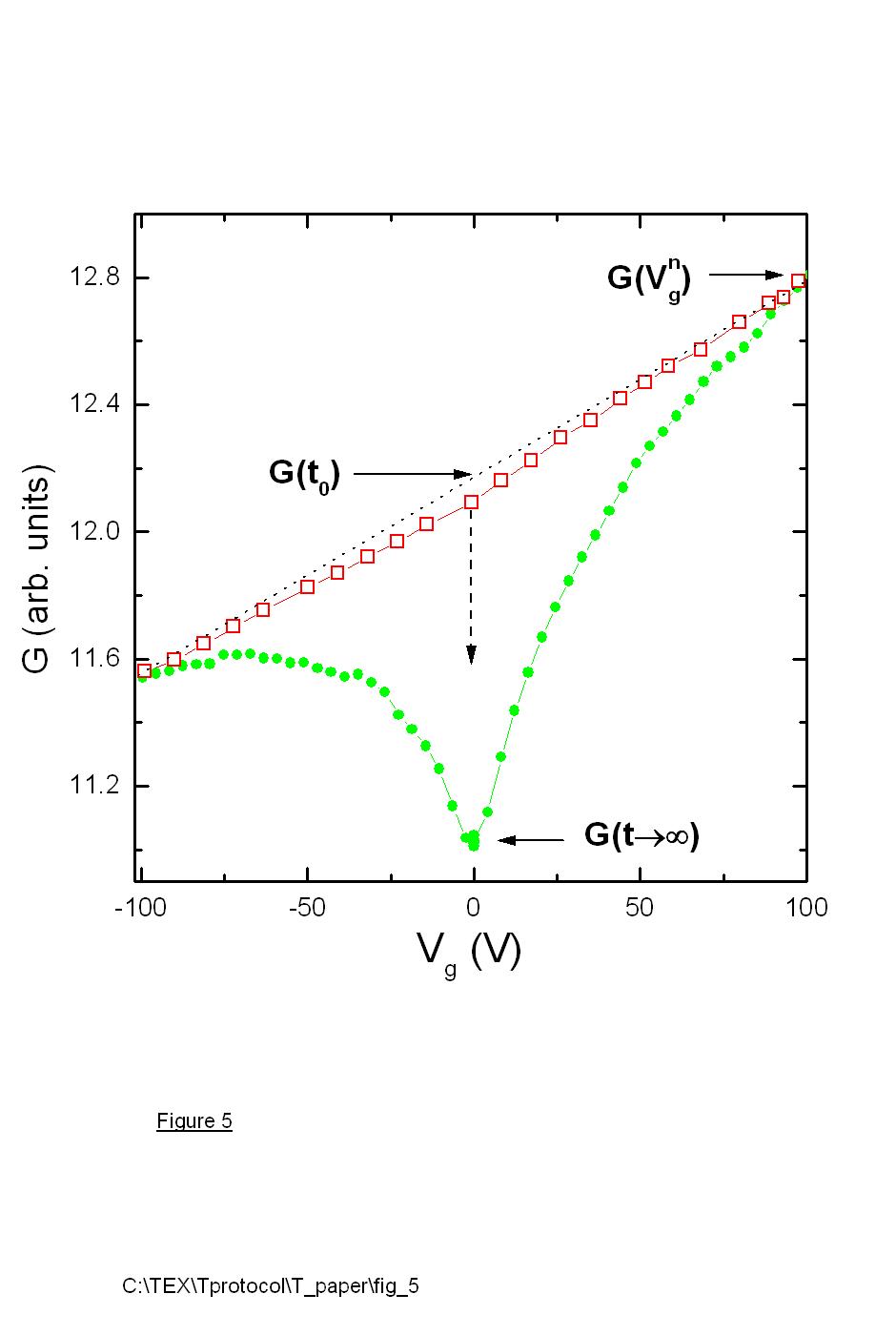}%
\caption{The memory cusp as recorded in a field effect plot $G(V_{g})$ (for
the sample in figures 1 and 2) short time (approximately $40\sec.)$ after the
quench from $100K$ (squares), and after 12 days (circles). The former trace
was taken by first sweeping $V_{g}$ to $+100V$ then recording $G(V_{g})$ by
continuously sweeping to $-100V$ with $8V/\sec$. sweep rate. The
`$t\rightarrow\infty$' trace was taken by sweeping from $V_{g}$ to either
extreme with $4V/\sec$. sweep rate. The dotted straight line connecting the
end points of $G(V_{g})$ is the field effect that is presumed to be recorded
if measured instantly after the quench.}%
\end{center}
\end{figure}
Immediately after a quench from high temperature $G(V_{g})$ shows just an
anti-symmetric field effect, which has essentially the same shape as the
equilibrium result. With time, $G(V_{g}=0)$ goes down along the dashed arrow
in figure 5. This is actually the process of the logarithmic conductance
relaxation (as observed in figure 6), unveiling the characteristic shape of a
cusp with a local minimum centered at $V_{g}^{o}=0.$ A model accounting for
such dynamics has been proposed by Yu \cite{18}. Note that the anti-symmetric
part of $G(V_{g})$ does not depend on the waiting time nor on the sweep rate
of $V_{g}$ \cite{13}. The contribution of the equilibrium field effect to $G$
when $V_{g}$ is switched to $V_{g}^{n}=100V$ from $V_{g}^{o}=0V$ is $\delta
G_{eq}$ and can be estimated from figure 5 as $\delta G_{eq}=\frac{G(V_{g}%
^{n})-G(-V_{g}^{n})}{2}.$ (Alternatively, one may use $\delta G_{eq}%
=G(V_{g}^{n})-G(t_{0})$ from the $G(V_{g}=0,t)$ plot by extrapolating to the
value of the conductance at $t_{0}$ along the well defined $\log(t)$
dependence as in figure 6. The $\delta G_{eq}$ obtained by these two ways
coincide to within the experimental error). This value of $\delta G_{eq}$ is
drawn in figure 2b as a dashed line, and is in fair agreement with the
asymptotic value of $\Delta G(t/t_{w}).$

The quality of data collapse is almost as good in the T-protocol as in the
gate-protocol except for the region $t\gg t_{w}$ where the scatter is more
noticeable. The main source of error in T-protocol is presumably the
uncertainty in the extrapolated function of $G_{b},$ which naturally is more
pronounced for large $t/t_{w}.$ The best one can do is to correct for the
effect of variations in the bath temperature during the relaxation process. In
the current case, temperature variations was apparently not the only reason
for the imperfect collapse for the $t\gg t_{w}$ regime. A probable cause is
the presence of some burst noise discussed above.%
\begin{figure}
[ptb]
\begin{center}
\includegraphics[
trim=0.000000in 4.052699in 0.000000in 1.871037in,
natheight=14.583300in,
natwidth=9.791400in,
height=2.8158in,
width=3.1808in
]%
{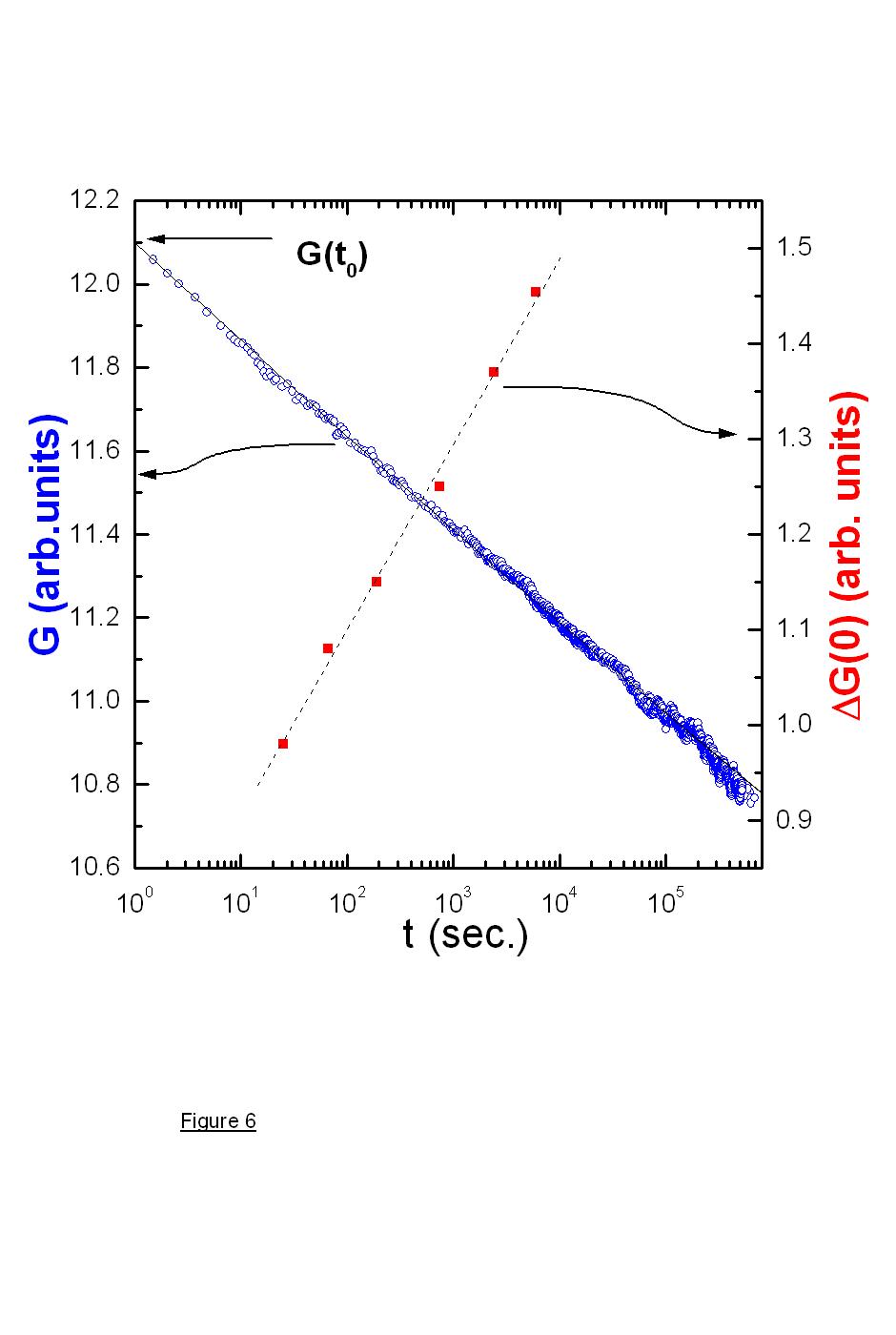}%
\caption{The conductance $G$ versus time following a quench from $100K$ to
$4.11K$ (circles) and the initial amlpitiude for the relaxation $\Delta
G(t_{0})$ following the switch from $V_{g}^{o}=0V$ to $V_{g}^{n}=100V$ for the
different waiting times in figure 2.}%
\end{center}
\end{figure}

While the T-protocol has a number of weaknesses as an experimental procedure,
it also has an advantage over the aging protocols previously used. In both,
gate-protocol and the stress-protocol the sample is initially at
\textit{equilibrium, }and $t_{w}$\textit{ }is the time it spends in an out of
equilibrium state. This requires that for each $t_{w}$ one has to wait for the
system to come to equilibrium, which, as may be inferred from figure 6 is a
long wait.\textit{ }In the T-protocol, on the other hand, the starting point
is an \textit{ergodic state,} so initializing the system for the next run is a
fast process. Also, this protocol seems the more natural way to probe `aging'
in the sense used in everyday's life; the system starts out "young", is then
"aged" for $t_{w},$ and the response $\Delta G$ due to the $V_{g}%
^{o}\rightarrow V_{g}^{n}$ change is a measure of how deeply the system has
"aged". This is judged by the initial amplitude and by the temporal dependence
of $\Delta G.$

As an experimental procedure however, the T-protocol is manifestly not as easy
as the two previously used protocols. The gate protocol, in particular, is
less 'invasive', much easier to control and implement, and therefore more reliable.

Aging in the most general form means that the response of the system (in our
case the excess conductance $\Delta G)$ is a function of both $t$ and
$t_{w}.~$It would appear that, to exhibit such a behavior, it is sufficient
that the system has a wide spectrum of relaxation times. Indeed, the response
due to the application of an exciting agent after $t_{w}$ will be composed of
only those components that have already relaxed. Therefore it is natural to
expect that the ensuing response will be both larger and include slower
components as $t_{w}$ increases. Clearly, these are the two ingredients that
characterize the aging phenomenon.

It is easy to show in the T-protocol that $\Delta G(0),$ the initial amplitude
of $\Delta G$ depends on $t_{w}~$in a unique way$.$ This comes naturally from
the basic properties of the electron glass, namely the logarithmic relaxation,
and the time evolution of the memory cusp. The $\log(t)$ relaxation of $G$
following a quench from a high temperature is shown in figure 6 for the sample
under study. Therefore, at time $t_{w}$ after the quench one finds:\newline%
$G(t_{w})=G(t_{0})-a_{o}\cdot\log(t_{w})$ (1)$\newline$where $t_{0}$ is the
time resolution of the experiment (typically, $t_{0}=1\sec.).$ When $V_{g}$
reaches $V_{g}^{n},$ the conductance increases to $G(V_{g})$ (see figures 1
and 5 for the definition of the various parameters). The initial amplitude for
the relaxation after the switch of $V_{g}$ is $\Delta G(0)\equiv
G(V_{g})-G(t_{w})$ and with (1):\newline$\Delta G(0)$=$G(V_{g})-G(t_{0}%
)+a_{o}\cdot\log(t_{w})$ (2)\newline The first two terms of the right hand
side of (2) include $\delta G_{eq}$ the equilibrium field effect, which as
discussed above is associated with the change with energy of the
\textit{thermodynamic} density of states. They also depend on the sweep rate
of $V_{g},$ the value of $T_{H},$ and the cooling rate involved in the quench.
The third term reflects the expected dependence on $t_{w}$; obviously the
longer the system ages the more susceptible it becomes to re-excitation
(`rejuvenation') i.e., a larger response will occur, simply because the
excitation affects only those components that have already relaxed. The
dependence of $\Delta G(0)$ on the waiting time for the six aging runs used in
figure 2 is depicted in figure 6.

While these considerations account for aging in its general form, the data
collapse upon scaling by $t/t_{w}$ illustrated in figures 2 and 4 cannot be
explained just by using the above assumptions, and a more fundamental
treatment appears to be needed. The heuristic model offered to account for
full-aging in the gate protocol \cite{19} is much harder to justify for the
T-protocol. The model involved two elements. The first is the logarithmic
relaxation law. This was shown to follow from the basic features of hopping
transport \cite{19} and it is supported by recent Monte-Carlo simulations
\cite{20}. The second was an assumption of some symmetry inherent to the
gate-protocol when $\delta V_{g}\equiv|V_{g}^{n}-V_{g}^{o}|$ is small enough.
It was later found that full aging is still observed even when this symmetry
is absent \cite{21}. The lack of symmetry is even more obvious in the
stress-aging-protocol which nonetheless show very good data collapse
\cite{11}. Moreover, the heuristic model could explain the data collapse only
in the regime $t\ll t_{w}~$where the logarithmic law is obeyed while the
master plots in all three protocols deviate from the logarithmic law for
$t\eqslantgtr0.2t_{w}.$ For the gate protocol the signal/noise is good enough
to confirm full aging behavior at least up to $t\approx2t_{w}$. For later
times $\Delta G$ reaches quickly the noise level as illustrated in figure 7
and we cannot be sure that there is still a perfect collapse. Nevertheless,
there seems to be a wide enough time interval over which full aging is
observed in the $t\gg t_{w}$ regime, and it is not describable by a
logarithmic time dependence. The right model for full aging in the electron
glass must be able to account for this regime as well as for the $t\ll t_{w}$ regime.

A model that gives full aging in both $t\ll t_{w}$ and $t\gg t_{w}$ regimes
has been suggested by Bouchaud and Dean \cite{22}. In their 'tree' model, the
two-time response is given as: $1-\frac{\sin(\pi x)}{\pi(1-x)}\left(  \frac
{t}{t+t_{w}}\right)  ^{y}$ and $\frac{\sin(\pi x)}{\pi x}\left(  \frac{t_{w}%
}{t+t_{w}}\right)  ^{x}$ for $t\ll t_{w}$ and $t\gg t_{w}$ respectively and
where $y\equiv1-x$. The data in figure 7 are plotted to allow comparison with
the $t\gg t_{w}$ regime. Note that to be consistent with the model, the value
of $x$ has to be about $0.87$ (c.f., figure 7). This is the average value of
the best-fit exponent found in eight aging experiments based on the
gate-protocol. The values for $x$ in these runs ranged between $0.85$ to
$0.91.$ This constrains the value of $y$ to be $0.09-0.15$ for the relaxation
in the $t\gg t_{w}$ regime. To reconcile the experimental results (figure 6)
with the model for the $t\ll t_{w}$ regime requires $y\leq0.0085,$ and given
the scatter in the data of figure 7 we cannot rule out agreement with the
model.%
\begin{figure}
[ptb]
\begin{center}
\includegraphics[
trim=0.000000in 3.116451in 0.000000in 1.871037in,
natheight=14.583300in,
natwidth=9.791400in,
height=3.1168in,
width=3.1808in
]%
{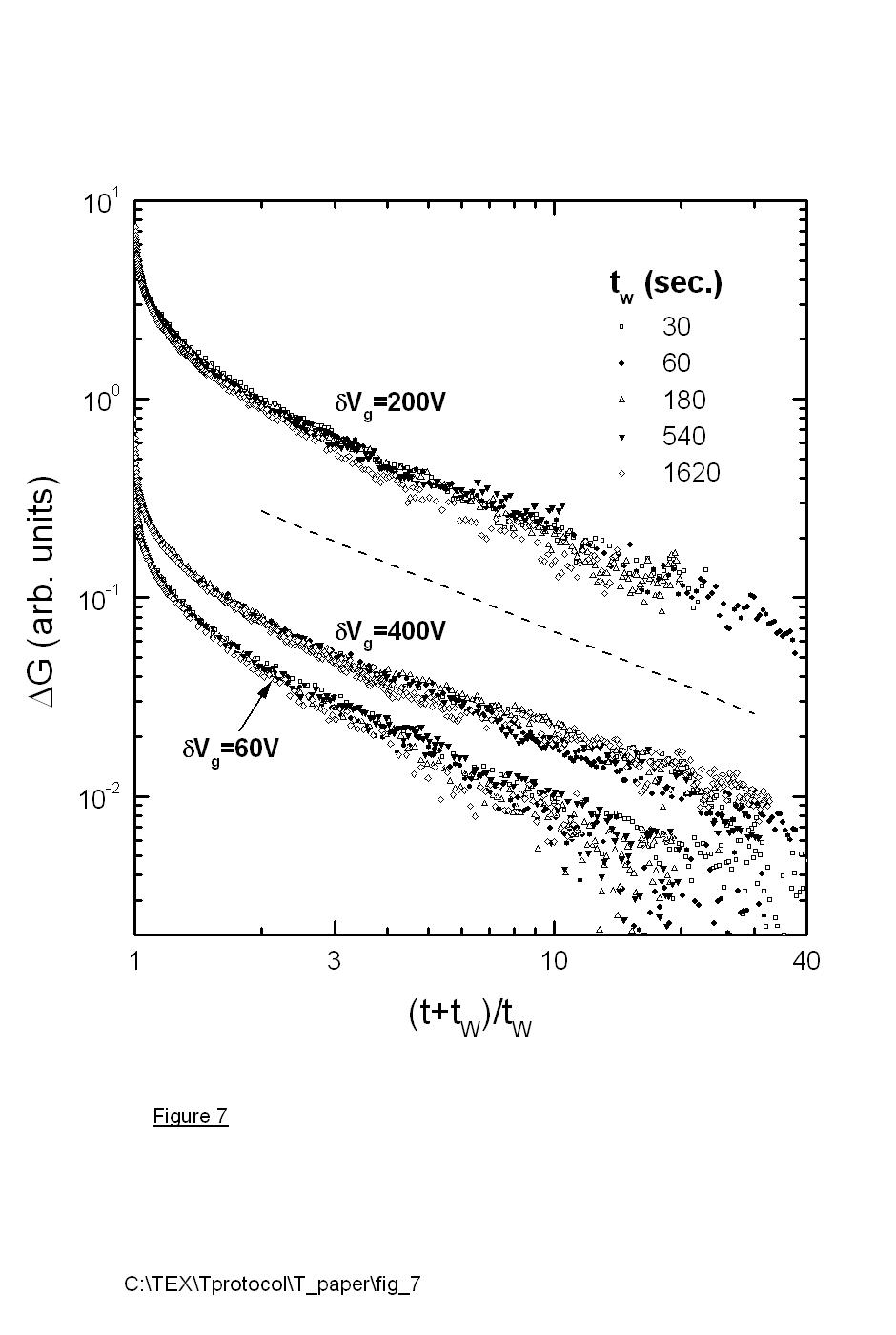}%
\caption{The excess conductance $\Delta G$ in a gate-protocol experiment as
function of the normalized time. The figure shows the results of three aging
series each with five values of waiting times. These were taken with two
different samples; The two bottom curves are data for a single $40.5M\Omega$
sample, where the two sets shown used different values of $\delta V_{g}$ as
indicated. The upper set was taken with a $51M\Omega$ sample with $\delta
V_{g}=200V.$ The latter data were shifted for clarity by multiplying $\Delta
G$ by $10$. Note that for $t\gg t_{w}$ the data is consistent with the
power-law behavior predicted by Bouchuad and Dean. The dashed line illustrates
a power-law with exponent $-0.87$ for comparison.}%
\end{center}
\end{figure}

This agreement however may be fortuitous. The Bouchaud and Dean model assumes
dynamics of a classical glass, where the barriers in phase space are crossed
by thermal activation. This is not appropriate for the case of a quantum glass
like the electron glass \cite{4}. In particular, the value of the exponent $x$
in the model is given by $T/T_{g}$ where $T_{g}$ is a glass temperature
\cite{11}. This gives the aging function a specific temperature dependence, in
contrast with the experiment \cite{9}. On the other hand the overall
resemblance of the model predictions to actual data suggest that the
underlying approach may have merit for real systems. This might encourage
researchers to look for a modification of the model that are more appropriate
for a glass with tunneling controlled dynamics rather than by thermal activation.

Using the gate-protocol it was shown before \cite{9} that the aging function
fits well a stretched exponent expression: $\Delta G(t,t_{w})\propto
\exp\left[  -\beta\left(  \frac{t}{t_{w}}\right)  ^{\alpha}\right]  $ with
$\beta\approx2.75$ and $\alpha\approx0.21.$ This expression, with strikingly
similar parameters $\alpha~$and~$\beta,$ were found to fit many aging
experiments on both crystalline and amorphous indium-oxide films. The
gate-protocol, recently applied to granular Al films, exhibited a near-perfect
full-aging, and seem to fit quite well to the stretched-exponent expression
above with $\beta\approx2.8$ and $\alpha\approx0.185~$\cite{10}$.$ Full-aging
behavior was observed in experiments based on the stress-protocol on
$In_{2}O_{3-x}$ films, and the results could be fitted by the
stretched-exponent with $\beta\approx2.32$ and $\alpha\approx0.215$ \cite{14}.
Finally, the aging function of the T-protocol could be fitted by a stretched
exponent with $\beta\approx2.65$ and $\alpha\approx0.22~$(after subtracting
0.58-0.6 that represents $\delta G_{eq}$ from the data in figure 2b). It is
important to note that these fits to the stretched exponent function should be
merely regarded as a convenient classification scheme; To see that the
stretched exponent cannot be the correct description of these experiments
suffice is to notice that for $t\ll t_{w}$ the function reduces to a power-law
dependence with exponent $\alpha.$ A value for $\alpha$ of the order of 0.2
may reasonably mimic a $\log(t)$ dependence over a limited range, which is why
this, in-principle discrepancy (c.f., figure 6), does not stand out in aging
experiments where $t_{w}$ was restricted to less than 4 decades. What the
similarity of parameters does tell us is that the aging function of several
systems must have a similar shape. This is not a trivial observation. The
studied systems; crystalline indium-oxide, several versions of amorphous
indium-oxides \cite{23}, and granular Al are totally different in terms of
microstructure, and they usually exhibit quite different conductance versus
temperature laws $G(T)$ \cite{24}. The feature that is common to all these
systems is that they are strongly localized, which is the pre-requisite to be
in the electron glass phase \cite{4}, and their $G(T)$ is consistent with
hopping transport mechanism \cite{24}. We suspect that the latter is an
important ingredient in bringing about the `unified' aging behavior; Namely,
it is the logarithmic relaxation inherent to the hopping mode of transport
\cite{13} that is common to all of the full-aging examples listed above.
Actually, the only difference between the results of different systems (and
different protocols), is the value at which the aging function $f\left(
t/t_{w}\right)  $ deviate from the \textit{generic} $\log(t)$ behavior. The
challenge is to find the physical scheme that controls this aspect of the
full-aging phenomenon.

The author acknowledges useful discussions with A. Efros and the hospitality
of the Physics department at Utah University where part of this work was
completed. This research was partially supported by a grant administered by
the US Israel Binational Science Foundation and by the Israeli Foundation for
Sciences and Humanities.


\begin{thebibliography}{99}
\bibitem {15}In particular, the time the system dwells at $T_{H}$ will play a
role in the ensuing relaxation.
\bibitem {23}Amorphous indium-oxide can be made with vastly different carrier
concentrations ranging between 10$^{18}$ to 10$^{22}cm^{-3}.$
\bibitem {24}These systems exhibit conductance versus temperature behavior
that is consistent with one of the variants of hopping conductivity, i.e.,
their $G(T)\propto\exp(T^{-p}),$ where $0.3\eqslantless$ $p\eqslantless$ $1$.
\end{thebibliography}
\end{document}